# The Impact of Content Commenting on User Continuance in Online Q&A Communities: An Affordance Perspective

Langtao Chen

Department of Business and Information Technology, Missouri University of Science and Technology, Rolla, Missouri, USA, chenla@mst.edu

## ABSTRACT

Online question-and-answer (Q&A) communities provide convenient and innovative ways for participants to share information and collaboratively solve problems with others. A growing challenge for those Q&A communities is to encourage and maintain ongoing user participation. From the perspective of motivational affordances, this study proposes a research framework to explain the effect of content commenting on user continuance behavior in online Q&A communities. The moderating role of participant's tenure in the relationship between content commenting and user continuance is also explored. Using a longitudinal panel dataset collected from a large online Q&A community, this research empirically tests the effect of content commenting on continued user participation in the Q&A community. The results show that both comment receipt and comment provisioning are important motivating factors for user continuance in the community. Specifically, received comments on questions submitted by a participant have a positive effect on the participant's continuance of posting questions, while answer comments both received and posted by a participant have positive impact on user continuance of posting answers in the community. In addition, tenure in the community is indeed found to have a significant negative moderating effect on the relationship between content commenting and user continuance. This research not only offers a more nuanced theoretical understanding of how content commenting affects continued user involvement and how participants' tenure in the community moderates the impact of content commenting, but also provides implications for improving user continuance in online Q&A communities.

## 1  Introduction

Online question-and-answer (Q&A) communities such as Reddit, Quora, Yahoo! Answers, and those hosted on the Stack Exchange network are growing as an important platform for online knowledge exchange that is not limited by spatial and temporal constraints. In 2019, Stack Exchange network had 3.1 million questions asked, 3.6 million answers submitted, 13.6 million comments posted, and 427.2 million monthly visits.[1] Participants actively engaging in online Q&A communities can easily solve their problems by locating experts or resources that are not possible, or at least not convenient, to access in offline settings. The success of online Q&A communities, which often lack economic incentives and institutional support, largely depends on the voluntary contribution of users. However, online communities, particularly those focusing on utilitarian knowledge exchange, often suffer from the "tragedy of commons" [20], a well-known dilemma which occurs when participants tend to free ride rather than spend their time and effort in building a public knowledge repository [11; 13; 42]. Consequently, an important theoretical and practical challenge for such online communities is to incentivize users' continued engagement [45].

To solve the under-engagement problem, modern social media platforms have adopted different types of affordances such as voting, text commenting, tagging, and feedbacks that can enrich user interactions and enhance usability and sociability of the platforms. This study explores content commenting as an important motivational affordance implemented in an online Q&A platform by evaluating its impact on users' continued involvement in the community. In a Q&A community that supports content commenting features,

---

participants can post text comments on the exchanged content including questions and answers. These comments posted in the community are not just a channel for communicating information regarding the content, but also an important design feature that can impact the perceived affordances of the online platform by the participating users. Extant literature generally supports that text comments are an important form of feedback in various online settings such as online marketplaces [40], online news communities [5], and online knowledge communities [11]. In the context of online Q&A communities, content commenting in the form of online texts can enhance social interactions in websites, thus ensuring sustainability and success of these online communities [11].

Furthermore, the length of membership (i.e., tenure) may impact user continuance behavior. Tenure has been identified as one important determinant of user contribution in online community literature [27; 30; 46]. It is generally treated as accumulated investment [43] or cognitive capital [46], which in turn promotes continued engagement. Although the current literature has intensively addressed the direct link between tenure and user participation in online communities, little attention has been paid to how different forms of online engagement change with respect to members' tenure in the community [36].

While previous research on online knowledge communities has addressed the impact of content commenting on knowledge contribution [11], there is limited research on how content commenting including both comment receipt and comment provisioning affects users' continued involvement in online Q&A communities, and in particular, how participants' tenure in the community interacts with the motivational affordances of content commenting. Thus, this research attempted to fill the gaps in the literature by answering the following questions:

*RQ1*: How does content commenting in the form of both comment receipt and comment provisioning impact user continuance of questioning and answering in online Q&A communities?

*RQ2*: How does the tenure of participants impact the relationship between content commenting and user continuance in online Q&A communities?

This research aims to make contributions by proposing a research model that explains how content commenting impacts user continuance and how participants' tenure plays a moderating role in the effect of content commenting on user continuance. A longitudinal panel dataset was collected from a large online Q&A community to empirically test the hypotheses. Results confirm the positive effects of comment receipt and provisioning on continued user engagement in the community. Moreover, participants' tenure moderates the impact of content commenting on user continuance. These findings enrich our understanding of how content commenting as a motivational affordance affects user behavior in online Q&A communities. In addition, this research provides practical implications for developers and managers to design their online platform features, mechanisms, and policies to encourage participants to engage more actively in online questioning and answering behaviors.

The rest of the paper is organized as follows. The next section outlines related research and proposes a research model and hypotheses. The third section introduces the research method, followed by empirical results in the fourth section. Next, theoretical and practical implications of this research are discussed. The last section concludes the research.

## 2  Theoretical Foundations and Hypotheses

As presented in Figure 1, the proposed research model incorporates a participant's comment receipt and provisioning to explain the participant's continuance of questioning and answering in online Q&A communities. Further, tenure of the participant is proposed to moderate the effects of comment receipt and provisioning on user continuance. The theoretical foundations and detailed hypotheses are provided in the following subsections.

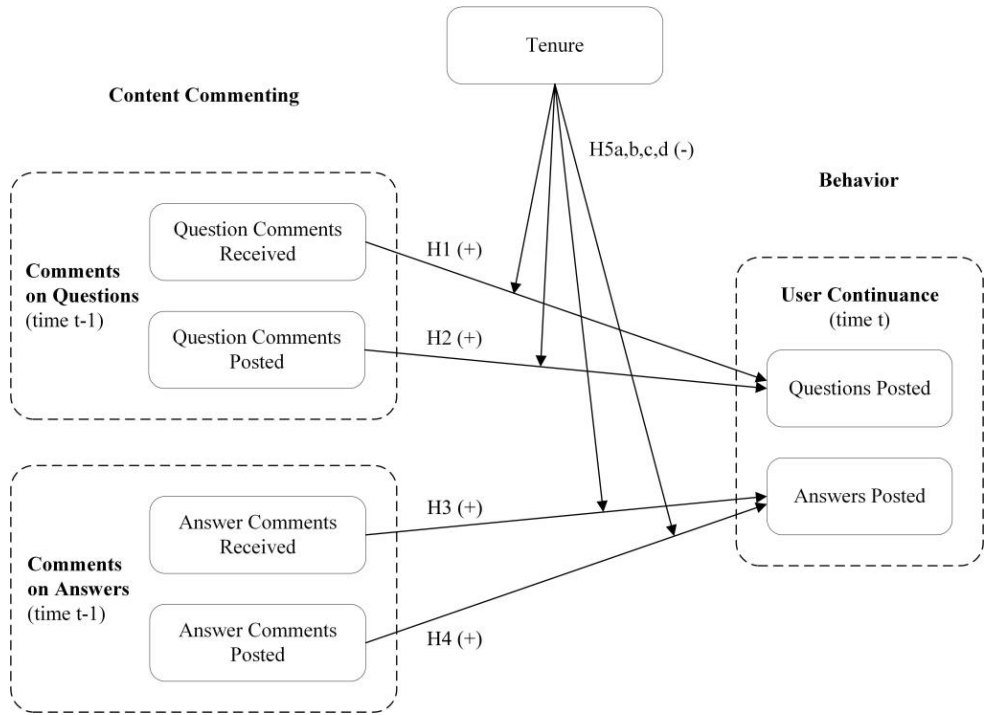

**Figure 1: Research model**

## 2.1 Commenting as motivational affordances

The term affordances generally refers to an object's actionable properties that can satisfy people's particular needs [17]. An object's affordances, both real and perceived, determine how it would be understood and used in a particular context [35]. In the setting of human-computer interaction, affordances are those properties of information system design features that invoke user motivations, interests, and enjoyments that further lead to interaction outcomes such as continued involvement and contribution [9; 11; 19]. Specifically, *Motivational affordances* are the system's properties that determine whether and how it can fulfill users' motivational needs [48]. The general design principle of human-computer interaction is that system features can and should be properly designed with high motivational affordances based on a set of psychological, social, cognitive, and emotional sources of motivation [48].

Social media platforms including Q&A sites have offered different types of motivational affordances such as voting, commenting, tagging, and feedback to better support social interactions [11; 38]. While previous research has addressed the effect of motivational affordances in enhancing user involvement in a set of information system design and management contexts such as wikis [1], digital health communities [2], and gamified design [25], insufficient attention has been paid to content commenting as an important motivational affordance in social media platforms including online Q&A communities [11]. This research takes a step forward in investigating the motivational affordances of content commenting in online Q&A communities.

## 2.2 Effects of content commenting on user continuance

In the setting of information systems, user continuance refers to the continuous use of a specific information system platform beyond its initial adoption [3]. User continuance is central to the success of online communities that rely heavily on the voluntary contribution of members, given that acquiring new members is often much more difficult than retaining existing ones. The situation would become even more challenging when the communities do not have well-designed features with high motivational affordances. The two major types of user continuance in online Q&A communities are to post questions and to answer other participants' questions over time. Such user continuance behaviors are, at least partially, determined by user experiences with a set of social and technical elements in the online communities.

According to social exchange theory [16; 18], social behavior is the result of social exchange or social interaction. In various online community settings, the current literature supports that social interaction promotes user participation [8; 11; 24]. As an important form of social interaction in online communities, content commenting allows participants to discuss content posted in the community, thus encouraging continued engagement in content building. The rationale behind this premise is that the amount of content commenting indicates the degree of social interaction of participants who receive and/or post comments as well as their extent of connectedness with others in the community [11]. Participants involving more deeply in content commenting can perceive a higher degree of belonging to the community [44]. It has been suggested by Oeldorf-Hirsch and Sundar [38] that comments promote a greater sense of influence and greater involvement for users sharing news stories on Facebook.

According to expectation-confirmation theory [39], expectation and perceived performance lead to user satisfaction which further influences user continuance intention [3]. Online Q&A community participants likewise go through a non-trivial decision process of expectation-confirmation. Specifically, participants involving more in online content commenting would perceive a higher level of usability and sociability of the online community platform and thus would more likely confirm their initial expectation that posting content is valuable to both the participants and the community as a whole. Such confirmed expectation would motivate participants to delve more deeply into continued content posting. Content commenting as a vital motivating affordance has been supported by empirical results in a set of online contexts. Brozowski et al. [4] found online comments to be associated with participants' continued contribution in an enterprise social media platform. Burke et al. [5] found that new members of Facebook receiving comments on their photos are more likely to continue posting photos. Chen et al. [11] found that receiving comments on users' answers motivates ongoing contribution in an online knowledge community.

This study considers asking and answering questions as two inseparable components of user continuance in online Q&A communities. While a sufficient volume of answers submitted to the communities ensures the utilitarian value of problem solving for participants in need, the sustainability of Q&A communities relies largely on a large volume of interesting and high-quality questions continuously posted to the communities. Moreover, rather than only focusing on receiving content comments from others [5; 11], this research also takes into account the provisioning of content comments by the focal participant, as receiving and provisioning content comments are part and parcel of meaningful social interaction in Q&A communities. This allows for unpacking the motivational mechanism of content commenting, thus providing opportunity for a more nuanced understanding of the motivational affordances of commenting. Based on the reasoning discussed above, receiving and posting comments on questions would enhance the feeling of social connectedness with others in the community, thus making question asking an enjoyable experience. Such enhanced feeling strengthens the perceived worth of question posting, which in turn would incentivize the focal participant involving in content commenting to continue to post questions in the community. Consequently, the following hypotheses are suggested:

*H1*: A participant receiving more question comments from the online Q&A community is likely to post more questions in the community.

*H2*: A participant posting more question comments is likely to post more questions in the online Q&A community.

Similarly, receiving and posting more comments on answers would enhance the perceived worth of answer posting and motivate a higher level of user continuance in posting answers in the community. Thus, the following hypotheses are proposed:

*H3*: A participant receiving more answer comments from the online Q&A community is likely to post more answers in the community.

**H4**: A participant posting more answer comments is likely to post more answers in the online Q&A community.

## 2.3 Moderating effects of tenure

In the setting of online communities, tenure refers to the length of time a participant has been involved in the community. As an important indicator of user experience, community tenure can shape a user's perception of social interaction and affect her/his behavior. There is conflicting evidence as to how tenure impacts user engagement in online communities. In the setting of an organizational bulletin board system, Wasko and Faraj [46] found a positive relationship between users' tenure and the volume of knowledge contribution. However, Khansa et al. [27] found that members with longer tenure contribute fewer questions and answers in the Yahoo! Answers website. In the setting of an online photo-sharing community, Nov et al. [36] found that participants' tenure in the community moderates the relationships between a set of motivations and user participation. Moreover, Ma and Agarwal's [30] study found tenure to be not significantly associated with user engagement. These inconsistent empirical results suggest that tenure likely interacts with other motivating factors. However, no sufficient attention has been paid to the moderating or interacting role of tenure in current online community literature.

To fill the gaps in the literature, this research aims to explore the moderating role of tenure in the relationship between content commenting and user continuance. Such a moderating role can be understood from two theoretical lenses. The first theoretical framework is the social capital theory [6; 33], which posits that collective action is affected by social capital in three forms including: (1) structural capital (overall structural links or connections between actors), (2) relational capital (individual relationships built through interaction), and (3) cognitive capital (resources enabling shared representations, interpretations, and meaningful understanding of knowledge). An indicator of cognitive capital, tenure in a community is accumulated through social interaction over time as participants develop their experiences and learn skills and norms of practice [46]. Thus, participants with longer tenure are more likely to have already accumulated more cognitive capital, which drives repeated user engagement in the community [37; 46]. With greater knowledge about the online community mechanisms and higher confidence in their own behavior, those senior members are less influenced by motivational affordances of content commenting. In contrast, participants with shorter tenure are more likely to be motivated by content commenting affordances as they lack requisite cognitive capital.

Secondly, from the perspective of information systems habits [28; 29; 41], continued use of an information system (e.g., an online Q&A community) over time is largely a habit rather than a conscious behavior. Although initial adoption of an information system involves careful evaluation of various factors such as ease of use and usefulness of the information system [14], repeated use of a system would become an automatic process without deliberate assessment of benefit and cost in a familiar context. Similarly in the setting of online Q&A communities, user continuance likely becomes a habitual and automatic behavior over time for participants with longer tenure. With this tendency toward automatic decision making, senior participants are less susceptible to the motivational affordances experienced during content commenting. Thus, the relationship between content commenting and user continuance is expected to be less evident for participants with longer tenure.

Both theoretical lenses discussed above lead to the negative moderating effect of tenure on the relationship between content commenting and user engagement. Specifically, the following relationships are expected:

**H5a**: A participant's tenure moderates the effect of received question comments on question posting in such a way that the effect is less positive when the participant has a longer tenure in the online Q&A community.

**H5b**: A participant's tenure moderates the effect of posted question comments on question posting in such a way that the effect is less positive when the participant has a longer tenure in the online Q&A community.

**H5c**: A participant's tenure moderates the effect of received answer comments on answer posting in such a way that the effect is less positive when the participant has a longer tenure in the online Q&A community.

***H5d***: A participant's tenure moderates the effect of posted answer comments on answer posting in such a way that the effect is less positive when the participant has a longer tenure in the online Q&A community.

## 3 Methods

### 3.1 Research setting and data collection

A field study was conducted to test the proposed research model. The research setting is superuser.com, a popular community-based Q&A website hosted on the Stack Exchange network for computer enthusiasts and power users. In this community, participants can register for free to post questions on specific topics in information technology or submit answers to other users' questions. The community supports a content commenting feature which allows participants to discuss particular questions or answers. Content commenting is the major channel of communication for collaborative problem solving in the community. For example, a participant can submit a comment on a question to ask the question initiator for more information regarding the problem. A comment on an answer may provide suggestions for the contributor to improve or refine her/his answer. In addition, participants can submit positive or negative votes on the usefulness of the questions and answers submitted by others to the community. The community also grants badges of three levels (gold, silver, and bronze) and assigns reputation scores to participants according to certain criteria related to their content contribution.

A rich dataset of user participation activities from August 2008 to December 2019 was collected. Figure 2 shows the monthly activities in the community. The whole dataset contains 419 thousand questions, 605 thousand answers, and 1.5 million comments. On average, each participant posted 1.42 questions, 2.05 answers, and 4.95 comments. Then a monthly panel dataset was created to test the hypotheses. Data before July 2009 were excluded from the analysis to alleviate the potential confounding effect due to the initialization of the community. The last month (December 2019) contained incomplete data and thus was also dropped.

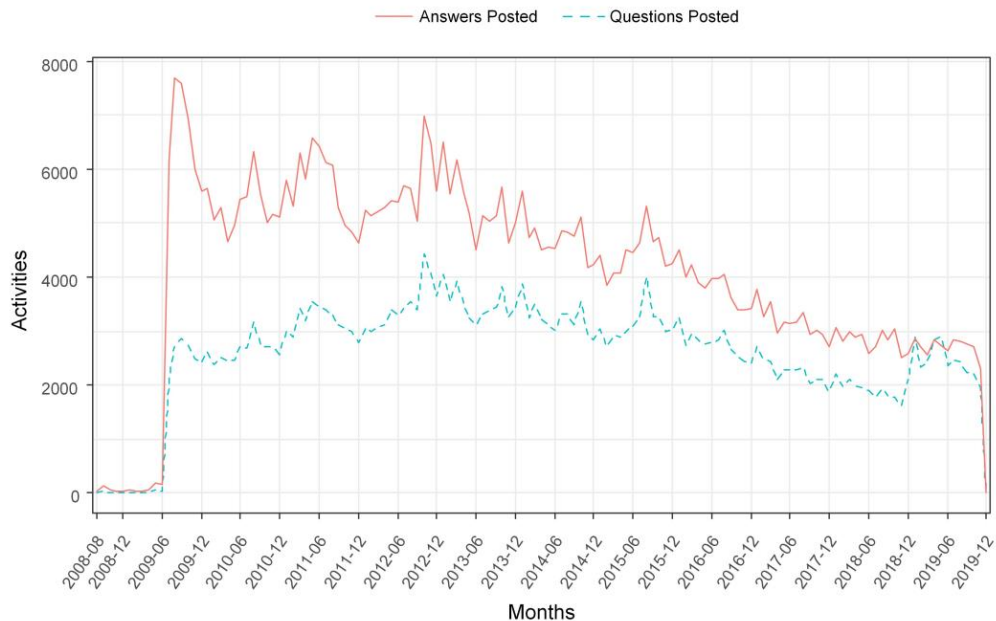

**Figure 2: Monthly activities in the Q&A community**

The longitudinal nature of the proposed research model, coupled with the fixed-effects Poisson regression model used for data analysis, requires a user to have observations at least in two months and post at least one question or answer. After refined to satisfy this requirement, the final unbalanced panel dataset contains 995,098 observations and 43,051 participants across 125 months (July 2009 through November 2019).

## 3.2 Variables

Table 1 presents the operationalized definition of variables used in this research. The dependent variables are questions and answers posted by a participant in a given month. The main independent variables include question comments received, question comments posted, answer comments received, answer comments posted by the participant and the tenure of the focal participant in a month. A number of factors that could be rival explanations of user continuance are controlled for. These control variables include: (1) the number of badges granted to the participant at each of the three levels including gold, silver, and bronze; (2) positive and negative votes the participant received for her/his questions posted to the community; (3) positive and negative votes the participant received for her/his answers posted to the community; (4) the reputation score of the participant at the beginning of a month; and (5) the number of answers the participant receives from the community. Table 2 shows the descriptive statistics and correlation coefficients of the variables.

## 3.3 Model specification and estimation

Following previous studies on online communities where complete information is often difficult to obtain [e.g., 10; 11; 26], this study applied fixed-effects panel regression methods to empirically analyze the dataset. Given that the two dependent variables are count variables, Poisson regression methods are suitable to estimate the effects of independent variables [7; 21]. Let $Question_{i,t}$ and $Answer_{i,t}$ denote the number of questions and answered posted by user $i$ ($i = 1, 2, \ldots, N$) in month $t$ ($t = 1, 2, \ldots, T$) respectively, the fixed-effects Poisson models are specified as follows:

$$Question_{i,t} = \exp \left( \begin{array}{l} \log(QCRec_{i,t-1})\beta_1 + \log(QCPos_{i,t-1})\beta_2 + Tenure_{i,t-1}\beta_3 + \\ \log(QCRec_{i,t-1}) \times Tenure_{i,t-1}\beta_4 + \log(QCPos_{i,t-1}) \times Tenure_{i,t-1}\beta_5 + \\ \log(ACRec_{i,t-1})\beta_6 + \log(ACPos_{i,t-1})\beta_7 + W_{i,t-1}\theta + month_t + \mu_i \end{array} \right) + \varepsilon_{i,t} \quad (1)$$

$$Answer_{i,t} = \exp \left( \begin{array}{l} \log(ACRec_{i,t-1})\beta_1 + \log(ACPos_{i,t-1})\beta_2 + Tenure_{i,t-1}\beta_3 + \\ \log(ACRec_{i,t-1}) \times Tenure_{i,t-1}\beta_4 + \log(ACPos_{i,t-1}) \times Tenure_{i,t-1}\beta_5 + \\ \log(QCRec_{i,t-1})\beta_6 + \log(QCPos_{i,t-1})\beta_7 + W_{i,t-1}\theta + month_t + \mu_i \end{array} \right) + \varepsilon_{i,t} \quad (2)$$

where $QCRec_{i,t-1}$ and $QCPos_{i,t-1}$ denote the number of question comments received and posted by user $i$ in month $t$-1 respectively; $ACRec_{i,t-1}$ and $ACPos_{i,t-1}$ are received and posted answer comments by user $i$ in month $t$-1 respectively; $Tenure_{i,t-1}$ is the tenure of user $i$ at the beginning of month $t$-1; $W_{i,t-1}$ represents other control variables; $month_t$ is a set of monthly dummies that control for potential time-specific effects; $\mu_i$ is the multiplicative participant-level fixed effects (or unobserved individual heterogeneity); $\varepsilon_{i,t}$ denotes the error term; $\beta$ and $\theta$ are the regression parameters to estimate. In addition to $W_{i,t-1}$, answers posted and questions posted in month $t$-1 are also controlled for in the question posting and answer posting models respectively. Hausman's [22] specification test supported the choice of fixed-effect Poisson models over random-effect Poisson models ($\chi^2 = 2501.14$ and 31806.91 for question posting and answer posting models respectively, p-values < 0.001).

**Table 1: Description of variables.**

| Variable type | Variable | Description |
|---|---|---|
| Dependent variables | Questions posted | The number of questions posted by a participant in a given month. |
| | Answers posted | The number of answers posted by the participant in a given month. |
| Independent variables | Question comments received | The number of comments that the participant receives from other users in a given month for questions submitted by the participant. |
| | Question comments posted | The number of comments that the participant posts for questions in a given month . |
| | Answer comments received | The number of comments that the participant receives from other users in a given month for answers submitted by the participant. |
| | Answer comments posted | The number of comments that the participant posts for answers in a given month. |
| | Tenure | The cumulative number of months that the participant has participated in the community. |
| Control variables | Gold badges | The number of gold badges granted to the participant by the community in a given month. |
| | Silver badges | The number of silver badges granted to the participant by the community in a given month. |
| | Bronze badges | The number of bronze badges granted to the participant by the community in a given month. |
| | Positive votes for questions | The number of positive votes (up votes) the participant receives in a given month for questions submitted by the participant. |
| | Negative votes for questions | The number of negative votes (down votes) the participant receives in a given month for questions submitted by the participant. |
| | Positive votes for answers | The number of positive votes (up votes) the participant receives in a given month for answers submitted by the participant. |
| | Negative votes for answers | The number of negative votes (down votes) the participant receives in a given month for answers submitted by the participant. |
| | Reputation | The reputation score of the participant in the community at the begining of a given month. |
| | Answers received | The number of answers the participant receives from the community in a given month. |

**Table 2: Descriptive statistics and correlation coefficients (N = 995,098).**

| Variables | Mean | S.D. | 1 | 2 | 3 | 4 | 5 | 6 | 7 | 8 | 9 | 10 | 11 | 12 | 13 | 14 | 15 |
|---|---|---|---|---|---|---|---|---|---|---|---|---|---|---|---|---|---|
| Questions posted | 0.13 | 0.61 | | | | | | | | | | | | | | | |
| Answers posted | 0.35 | 3.03 | 0.11 | | | | | | | | | | | | | | |
| Question comments received | 0.26 | 1.04 | 0.27 | 0.04 | | | | | | | | | | | | | |
| Question comments posted | 0.61 | 6.35 | 0.05 | 0.38 | 0.11 | | | | | | | | | | | | |
| Answer comments received | 0.41 | 3.3 | 0.05 | 0.69 | 0.05 | 0.48 | | | | | | | | | | | |
| Answer comments posted | 0.56 | 3.42 | 0.14 | 0.58 | 0.17 | 0.68 | 0.79 | | | | | | | | | | |
| Tenure | 0.44 | 0.89 | 0.10 | 0.07 | 0.17 | 0.06 | 0.07 | 0.10 | | | | | | | | | |
| Gold badges | 0.03 | 0.18 | 0.01 | 0.03 | 0.01 | 0.05 | 0.05 | 0.05 | -0.03 | | | | | | | | |
| Silver badges | 0.17 | 0.46 | 0.02 | 0.14 | 0.02 | 0.15 | 0.21 | 0.18 | -0.03 | 0.07 | | | | | | | |
| Bronze badges | 0.45 | 0.89 | 0.17 | 0.21 | 0.24 | 0.16 | 0.28 | 0.29 | 0.12 | 0.01 | 0.10 | | | | | | |
| Positive votes for questions | 0.76 | 2.81 | 0.15 | 0.03 | 0.28 | 0.04 | 0.04 | 0.10 | 0.03 | 0.11 | 0.15 | 0.18 | | | | | |
| Negative votes for questions | 0.03 | 0.25 | 0.14 | 0.02 | 0.35 | 0.06 | 0.03 | 0.10 | 0.07 | 0.02 | 0.03 | 0.13 | 0.24 | | | | |
| Positive votes for answers | 1.37 | 8.04 | 0.03 | 0.56 | 0.03 | 0.43 | 0.76 | 0.62 | 0.05 | 0.09 | 0.32 | 0.27 | 0.06 | 0.02 | | | |
| Negative votes for answers | 0.04 | 0.29 | 0.03 | 0.34 | 0.03 | 0.28 | 0.50 | 0.40 | 0.04 | 0.03 | 0.12 | 0.17 | 0.03 | 0.04 | 0.40 | | |
| Reputation | 303.03 | 893.01 | -0.06 | -0.03 | -0.07 | -0.03 | -0.02 | -0.06 | -0.11 | 0.04 | 0.07 | -0.09 | 0.09 | -0.02 | 0.09 | 0.02 | |
| Answers received | 0.36 | 1.22 | 0.41 | 0.06 | 0.51 | 0.07 | 0.07 | 0.23 | 0.18 | 0.01 | 0.03 | 0.28 | 0.36 | 0.28 | 0.05 | 0.05 | -0.07 |

# 4 Results

Table 3 presents the effect of content commenting on participants' continuance of posting questions. Column 1 only includes control variables. Column 2 adds primary independent variables, followed by column 3 which further includes monthly dummies to control for potential time effects. Column 4 includes interaction terms. Chi-square goodness fit test supports that data fit the fixed-effects Poisson models well (Wald $\chi^2$ ranges from 12419 to 14237, p-values < 0.001). Robust standard errors clustered at individual participant level are used to deal with potential heteroscedasticity, autocorrelation, and overdispersion of the data [23; 47]. The main effects are shown in column 3 of Table 3. The results confirm that receiving question comments from others has a positive effect on a user's continuance of posting questions in the community (β = 0.083, p-value < 0.001). Thus, hypothesis H1 is supported. However, the effect of posting question comments on user continuance of posting questions is not statistically significant (β = 0.009, p-value > 0.1). Therefore, hypothesis H2 is not supported. In addition, a participant's tenure in the community has a significant positive effect on the participant's continuance of posting questions (β = 0.136, p-value < 0.001). As shown in column 4 of Table 3, the two interaction terms both have significant negative effects (β = -0.073 and -1.169 respectively, p-values < 0.001), supporting hypotheses H5a and H5b that the positive effects of question comment receipt and provisioning become less positive for participants with longer tenure.

**Table 3: Effect of content commenting on question posting.**

|  | (1) | (2) | (3) | (4) |
|---|---|---|---|---|
| ***Primary variables*** |  |  |  |  |
| Question comments received (log) |  | 0.083*** | 0.083*** | 0.137*** |
|  |  | (0.011) | (0.011) | (0.014) |
| Question comments posted (log) |  | 0.008 | 0.009 | 0.722*** |
|  |  | (0.013) | (0.013) | (0.203) |
| Tenure |  | 0.136*** | 0.136*** | 0.008 |
|  |  | (0.005) | (0.005) | (0.038) |
| ***Interaction effects*** |  |  |  |  |
| Question comments received (log) × tenure |  |  |  | -0.073*** |
|  |  |  |  | (0.010) |
| Question comments posted (log) × tenure |  |  |  | -1.169*** |
|  |  |  |  | (0.329) |
| ***Control variables*** |  |  |  |  |
| Answer comments received (log) | -0.008 | 0.001 | 0.000 | 0.004 |
|  | (0.017) | (0.016) | (0.016) | (0.016) |
| Answer comments posted (log) | 0.118*** | 0.119*** | 0.118*** | 0.117*** |
|  | (0.011) | (0.011) | (0.011) | (0.011) |
| Gold badges (log) | -0.402*** | -0.406*** | -0.406*** | -0.404*** |
|  | (0.039) | (0.040) | (0.040) | (0.040) |
| Silver badges (log) | -0.348*** | -0.341*** | -0.340*** | -0.338*** |
|  | (0.017) | (0.017) | (0.017) | (0.017) |
| Bronze badges (log) | 0.378*** | 0.378*** | 0.374*** | 0.374*** |
|  | (0.011) | (0.011) | (0.011) | (0.011) |
| Positive votes for questions (log) | -0.142*** | -0.150*** | -0.150*** | -0.149*** |
|  | (0.009) | (0.009) | (0.009) | (0.009) |
| Negative votes for questions (log) | -0.022 | -0.057* | -0.057* | -0.057* |
|  | (0.027) | (0.027) | (0.027) | (0.027) |
| Positive votes for answers (log) | -0.077*** | -0.069*** | -0.068*** | -0.067*** |

|                                   | (0.013)      | (0.013)      | (0.013)      | (0.013)      |
|-----------------------------------|--------------|--------------|--------------|--------------|
| Negative votes for answers (log)  | -0.082**     | -0.076**     | -0.075*      | -0.074*      |
|                                   | (0.030)      | (0.029)      | (0.029)      | (0.029)      |
| Reputation/1000                   | -1.820***    | -1.785***    | -1.783***    | -1.769***    |
|                                   | (0.101)      | (0.098)      | (0.098)      | (0.097)      |
| Answers posted (log)              | 0.298***     | 0.269***     | 0.269***     | 0.266***     |
|                                   | (0.018)      | (0.017)      | (0.017)      | (0.017)      |
| Answers received (log)            | 0.512***     | 0.436***     | 0.436***     | 0.432***     |
|                                   | (0.012)      | (0.012)      | (0.012)      | (0.012)      |
| Monthly dummies                   | No           | No           | Yes          | Yes          |
| Log-likelihood                    | -265059      | -263890      | -263714      | -263645      |
| Wald $\chi^2$                     | 12419        | 13499        | 14124        | 14237        |

Notes: (1) Fixed-effects Poisson regression for models 1 through 4;
(2) *** p<0.001, ** p<0.01, * p<0.05, † p<0.1;
(3) Robust standard errors in parentheses.

In a similar way, Table 4 presents the estimation results of the answer posting model. The main results, as shown in column 3 of Table 4, confirm the positive effects of both answer comment receipt and answer comment provisioning on users' continuance of posting answers in the community ($\beta = 0.376$ and 0.247 for answer comment receipt and answer comment provisioning respectively, p-values < 0.001). Thus hypotheses 3 and 4 are both supported. Tenure in the community is found to have a significant positive effect on continuance of posting answers ($\beta = 0.211$, p-value < 0.001). In terms of the moderating role of tenure, column 4 of Table 4 shows that the effect of answer comment receipt on continued answer posting is less positive when the participant has a longer tenure in the online Q&A community ($\beta = -0.058$, p-value < 0.001). Thus, hypothesis H5c is supported. Similarly, the effect of answer comment receipt on continued answer posting is less positive for participants with longer tenure ($\beta = -1.373$, p-value < 0.001), providing support for hypothesis H5d. Consistent with the previous study by Wasko and Faraj [46], tenure of a community member is found to positively associate with continued participation behaviors.

**Table 4: Effects of content commenting on answer posting.**

|                                             | (1)        | (2)        | (3)        | (4)        |
|---------------------------------------------|------------|------------|------------|------------|
| *Primary variables*                         |            |            |            |            |
| Answer comments received (log)              |            | 0.375***   | 0.376***   | 0.412***   |
|                                             |            | (0.018)    | (0.018)    | (0.020)    |
| Answer comments posted (log)                |            | 0.247***   | 0.247***   | 1.091***   |
|                                             |            | (0.019)    | (0.019)    | (0.160)    |
| Tenure                                      |            | 0.211***   | 0.211***   | 0.023      |
|                                             |            | (0.008)    | (0.008)    | (0.037)    |
| *Interaction effects*                       |            |            |            |            |
| Answer comments received (log) × tenure     |            |            |            | -0.058***  |
|                                             |            |            |            | (0.014)    |
| Answer comments posted (log) × tenure       |            |            |            | -1.373***  |
|                                             |            |            |            | (0.258)    |
| *Control variables*                         |            |            |            |            |
| Question comments received (log)            | -0.155***  | -0.081***  | -0.079***  | -0.078***  |
|                                             | (0.021)    | (0.019)    | (0.019)    | (0.019)    |
| Question comments posted (log)              | 0.331***   | 0.092***   | 0.093***   | 0.092***   |
|                                             | (0.023)    | (0.019)    | (0.018)    | (0.018)    |
| Gold badges (log)                           | -0.142***  | -0.114***  | -0.114***  | -0.114***  |
|  | (0.037) | (0.031) | (0.031) | (0.031) |
| Silver badges (log) | -0.427*** | -0.324*** | -0.326*** | -0.326*** |
|  | (0.019) | (0.023) | (0.022) | (0.022) |
| Bronze badges (log) | 0.510*** | 0.435*** | 0.431*** | 0.431*** |
|  | (0.027) | (0.024) | (0.024) | (0.024) |
| Positive votes for questions (log) | -0.269*** | -0.218*** | -0.215*** | -0.215*** |
|  | (0.022) | (0.018) | (0.018) | (0.018) |
| Negative votes for questions (log) | -0.091† | -0.067 | -0.071 | -0.071 |
|  | (0.049) | (0.047) | (0.046) | (0.046) |
| Positive votes for answers (log) | 0.238*** | -0.076*** | -0.075*** | -0.075*** |
|  | (0.021) | (0.016) | (0.016) | (0.016) |
| Negative votes for answers (log) | 0.091*** | -0.054* | -0.054* | -0.054* |
|  | (0.027) | (0.025) | (0.025) | (0.025) |
| Reputation/1000 | -2.516*** | -2.150*** | -2.148*** | -2.141*** |
|  | (0.188) | (0.171) | (0.171) | (0.171) |
| Questions posted (log) | 0.664*** | 0.539*** | 0.538*** | 0.535*** |
|  | (0.036) | (0.034) | (0.034) | (0.034) |
| Answers received (log) | 0.012 | -0.073** | -0.076** | -0.075** |
|  | (0.026) | (0.024) | (0.024) | (0.024) |
| Monthly dummies | No | No | Yes | Yes |
| Log-likelihood | -442151 | -424768 | -424268 | -424169 |
| Wald $\chi^2$ | 6168 | 9069 | 10026 | 10059 |

Notes: (1) Fixed-effects Poisson regression for models 1 through 4;
(2) *** p<0.001, ** p<0.01, * p<0.05, † p<0.1;
(3) Robust standard errors in parentheses.

Figure 3 demonstrates the moderating effect of tenure on the relationship between content commenting and user continuance. The general pattern shown in Figure 3 (A), (B), and (C) is that the positive effect of content commenting becomes less positive as participants' tenure increases. Interestingly, as shown in Figure 3 (B), the relationship between question comment posting and continued question posting becomes slightly negative for participants with longer tenure (at the value of mean + S.D.). This provides an alternative explanation of why the overall relationship between question comment posting and continued question posting is not statistically significant (refer to Table 3 column 3). These results also suggest that it is imperative to integrate tenure into considering the motivational affordances of content commenting in the setting of online Q&A communities.

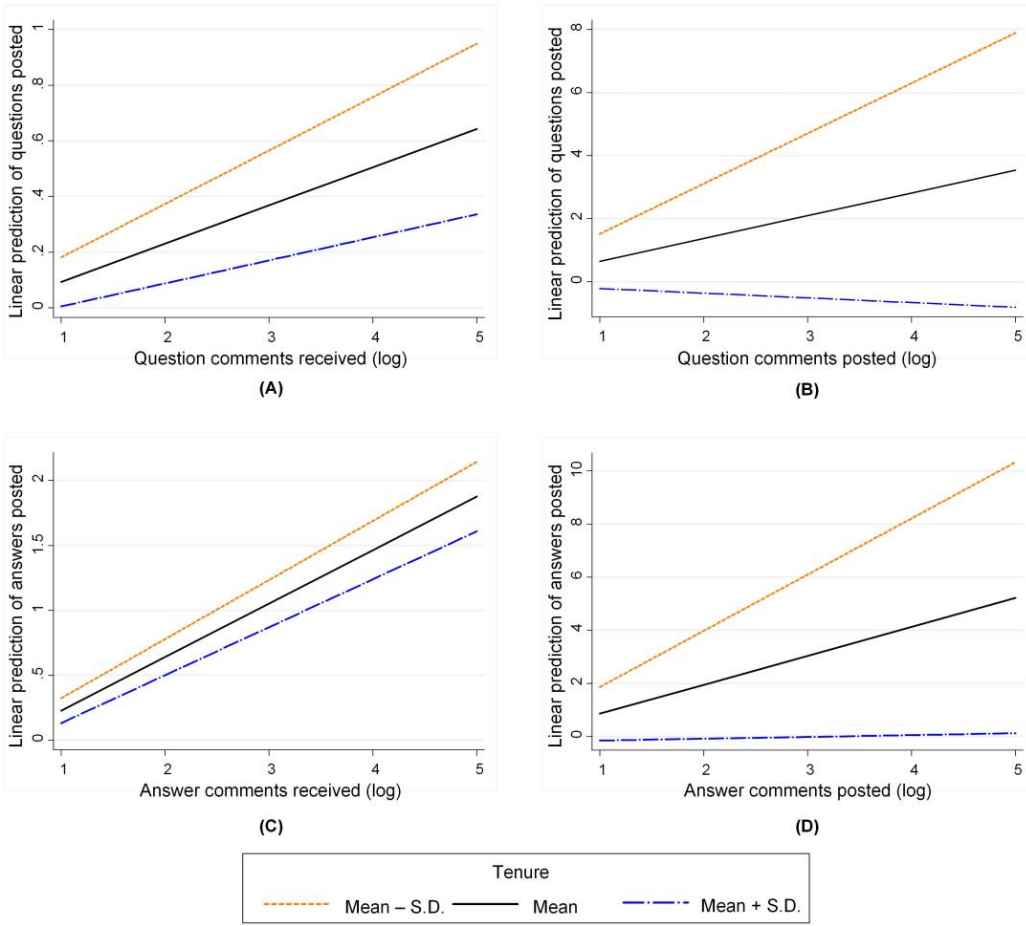

**Figure 3: Moderating effects of tenure**

# 5 Discussion

Online Q&A communities have grown as an important knowledge exchange platform for professional information sharing and collaborative problem solving. Not surprisingly, online Q&A communities have become an emerging research theme that potentially provides opportunities to significantly contribute to multiple research fields such as information systems, human-computer interaction, and other related fields. This research aims to answer the question of how content commenting affects user continuance as well as how participants' tenure plays a moderating role in the relationship between content commenting and user continuance. Content commenting is a fascinating online community feature, particularly in communities that focus on professional questioning and answering.

However, there lacks a theoretical framework that comprehensively models the impact of content commenting on continued question asking and answer provisioning in online Q&A communities, especially when the moderating role of participants' tenure is also taken into account. Few research efforts in the current literature have combined these elements in the same analysis that focuses on user continuance in online Q&A communities. Combining the perspective of motivational affordances and other theoretical bases, this research suggests that content commenting in online Q&A communities provides motivational affordances that help the communities to maintain continued user engagement. Furthermore, tenure in the community is found to negatively moderate the impact of content commenting on user continuance. Table 5 summarizes the results of hypothesis testing.

**Table 5: Summary of hypothesis testing.**

| No. | Hypothesis | Supported? |
|-----|-----------|------------|
| H1 | A participant receiving more question comments from the online Q&A community is likely to post more questions in the community. | Yes |
| H2 | A participant posting more question comments is likely to post more questions in the online Q&A community. | No |
| H3 | A participant receiving more answer comments from the online Q&A community is likely to post more answers in the community. | Yes |
| H4 | A participant posting more answer comments is likely to post more answers in the online Q&A community. | Yes |
| H5a | A participant's tenure moderates the effect of received question comments on question posting in such a way that the effect is less positive when the participant has a longer tenure in the online Q&A community. | Yes |
| H5b | A participant's tenure moderates the effect of posted question comments on question posting in such a way that the effect is less positive when the participant has a longer tenure in the online Q&A community. | Yes |
| H5c | A participant's tenure moderates the effect of received answer comments on answer posting in such a way that the effect is less positive when the participant has a longer tenure in the online Q&A community. | Yes |
| H5d | A participant's tenure moderates the effect of posted answer comments on answer posting in such a way that the effect is less positive when the participant has a longer tenure in the online Q&A community. | Yes |

## 5.1 Theoretical implications

Unlike a set of previous studies that only focus on answer provisioning [11; 30; 37; 46], this research contributes to the literature by considering both question asking and answer provisioning as two inseparable essential components of user continuance in online Q&A communities. Individually modeling question asking and answer provisioning helps to reveal the different motivating roles of content commenting in continuance of these two activities. Similarly, receipt and provisioning of content comments are separately assessed to evaluate whether they have similar effects on user continuance. This helps to unpack the motivational mechanism of content commenting, thus potentially providing a more nuanced understanding of what aspects of content commenting actually have motivational affordances.

Further, this research makes a distinctive contribution by incorporating both content commenting and participants' tenure to comprehensively understand user continuance behavior in utilitarian online Q&A communities. This study shows that content commenting on both questions and answers plays a vital role in users' continued involvement in online Q&A sites, extending previous work that only investigated the impact of content commenting on knowledge contribution [11]. Interestingly, receipt of question comments is found to have a significant positive effect on continued question posting behavior, while provisioning of question comments does not have a significant effect on continuance of question asking. This highlights the importance of investigating the moderating role of other important factors such as participants' tenure in order to fully understand the motivational affordances of design features in online communities.

Lastly, this research extends previous studies by finding that members with longer tenure tend to repeat their online questioning and answering behaviors over time, even when they experience a lower level of the affordance of content commenting. Users with different levels of knowledge or experience of online questioning and answering exhibit different behavioral patterns. Participants' tenure in the community indeed has a moderating effect on the receipt and provisioning of question and answer comments with regard to their relationships with continued user participation in the form of posting questions and answers to the community. With the consideration of the interaction effects of users' tenure and motivational affordances, this research helps to reconcile inconsistencies in previous studies [27; 30; 46] which suggested conflicting evidence regarding the effect of users' tenure. The confirmation of the moderating role of tenure also provides new thinking of users' tenure in other settings of online communities such as enterprise social media [4; 31; 32] and those focusing on health [10; 34], or other general contexts of information system design or human-computer interaction.

## 5.2 Practical implications

This research provides useful insights into designing motivational affordances for online Q&A communities that often suffer from the free-ride issue. Given that one important goal of online Q&A community designers and managers is to design and management an online platform that continues to support information sharing and collaborative problem solving, findings of this research suggest that these designers and managers should seriously consider content commenting features as important motivational affordances, with the careful evaluation of solutions and policies for users having different tenure of community experience. For a particular online community platform, designers and managers can use the insights of this research to guide their assessment of what commenting features invoke the preferred user participation behaviors and how participants' tenure affects the relationships. Such deliberate assessment would help the designers and managers to build an online community with optimal motivational affordances, thus preventing or at least reducing the destructive impact of the "tragedy of commons" [20].

Furthermore, the empirical results show that users' tenure negatively moderates the effects of commenting on user continuance in posting questions and answers to the community. This finding provides new thinking regarding the design and organization of online Q&A communities. As members with longer tenure are less motivated by the commenting mechanism of the community, other design features such as gamified ones (i.e., applying game design elements [15]) are the relatively more preferred mechanisms for retaining those senior members. This is particularly important for the survival and success of online Q&A communities, given that senior members with more knowledge and expertise are often the most valuable resources who lead the online social interaction and problem solving.

On the other hand, the positive effect of content commenting is more evident in members with shorter tenure. This suggests that online Q&A communities may need to adapt the commenting features according to members' tenure such that the junior members are more exposed to the commenting features and their questions or answers are ranked in a way that encourages more comments from other participants. Further, online reputation or reward systems could be fine-tuned to encourage content commenting for different user groups.

In sum, the findings of this research implied several opportunities for the online Q&A community developers and managers to enhance the platform features to more effectively incentivize continued user engagement and ultimately solve the problem of under-engagement.

## 5.3 Limitations and future research

The present research has several limitations. The findings in this research are specific to an online Q&A community that focuses on IT-related topics. Future research could analyze other online communities that have different user population, topics, or organizational structures.

Furthermore, this research only considers the receipt and provisioning of comments without actually analyzing the content of the comments. Future research can take into account the meanings embedded in those textual comments posted in online Q&A communities. Various linguistic cues embedded in the comments such as affective and informative signals [12] could be further evaluated in order to develop user continuance theory at more granular levels.

Moreover, an opportunity for future research is to use other field studies such as experimental studies or qualitative case studies to uncover the detailed mechanism of content commenting on user engagement by considering other important theoretical constructs. Such work may triangulate the findings of this study or better explain the motivational affordances of content commenting in online Q&A communities.

# 6 Conclusion

Online Q&A communities have become an important platform that helps users to solve complex questions which are usually hard to find solutions directly through information search or offline social network. As many online Q&A communities have been confronting the under-engagement challenge, the question of how to motivate user participation in online Q&A communities with high motivational affordances has inspired a variety of studies. Adopting the perspective of motivational affordances, this research enriches current understandings of the motivating role of content commenting in online Q&A communities, especially when tenure of participants is also taken into account. Based on a 125-month longitudinal panel dataset collected from a large online Q&A community, the analysis confirms the positive effect of content commenting on user continuance as well as the moderating role of users' tenure. The findings provide theoretical and practical insights into stimulating continued questioning and answering behaviors in online Q&A communities.